\documentclass{mn2e}
\usepackage{psfig}

\newcommand{\etal}{et~al.}

\newcommand{\ionhy}{H{\sc ii}}

\newcommand{\water}{$\mbox{H}_{2}\mbox{O}$}

\newcommand{\transa}{$7_{2}\mbox{-}6_{3}\mbox{~A}^{-}$}
\newcommand{\transe}{$6_{-2}\mbox{-}7_{-1}\mbox{~E}$}

\newcommand{\degrees}{$^\circ$}
\newcommand{\kms}{$\mbox{km~s}^{-1}$}
\newcommand{\cc}{\mbox{$\,\rm cm^{-3}$}}
\newcommand{\ccs}{\mbox{$\,\rm cm^{-3} s$}}
\newcommand{\nH}{\mbox{$n_{\rm H}$}}
\newcommand{\scdm}{\mbox{$N_{\rm M}/\Delta V$}}
\newcommand{\Td}{\mbox{$T_{\rm d}$}}
\newcommand{\Tk}{\mbox{$T_{\rm k}$}}

\newcommand{\beam}{\mbox{$\epsilon^{-1}$}}
\newcommand{\HII}{H\,{\sc ii}}
\newcommand{\vt}{\mbox{$v_{\rm t}$}}

\newcommand{\specdfig}[2]        
{
  \begin{center}
    \begin{minipage}[t]{0.45\textwidth}
        \psfig{file=#1.ps,height=0.75\textwidth,angle=270}
    \end{minipage}
    \hfill
    \begin{minipage}[t]{0.45\textwidth}
        \psfig{file=#2.ps,height=0.75\textwidth,angle=270}
    \end{minipage}
  \end{center}
}

\newcommand{\specsfig}[1]        
{
  \begin{center}
    \begin{minipage}[t]{0.45\textwidth}
        \psfig{file=#1.ps,height=0.75\textwidth,angle=270}
    \end{minipage}
  \end{center}
}
\newcommand{\speclfig}[2]        
{
  #1
  \begin{center}
    \begin{minipage}[t]{0.45\textwidth}
      \psfig{file=#2.ps,height=0.75\textwidth,angle=270}
    \end{minipage}
  \end{center}
}

\begin{document}

\title[85.5- and 86.6-GHz methanol maser emission] 
{A search for 85.5- and 86.6-GHz methanol maser emission}

\author[Ellingsen \etal\/]{S.P. Ellingsen$^1$, D.M. Cragg$^2$, 
  V. Minier$^3$, E. Muller$^4$, P.D. Godfrey$^2$\\
$^1$ School of Mathematics and Physics, University of Tasmania, 
     Private Bag 21, Hobart, Tasmania 7001, Australia;\\  
     Simon.Ellingsen@utas.edu.au\\
$^2$ School of Chemistry, Building 23, Monash University, Victoria 3800, 
     Australia;\\
     Dinah.Cragg@sci.monash.edu.au, Peter.Godfrey@sci.monash.edu.au\\
$^3$ Department of Astrophysics and Optics, School of Physics,
     University of New South Wales, NSW 2052, Australia;\\
     vminier@bat.phys.unsw.edu.au\\
$^4$ Department of Engineering Physics, University of Wollongong,
     Northfields Avenue, NSW 2522, Australia;}

\maketitle

\begin{abstract}
  
  We have used the Australia Telescope National Facility Mopra 22m
  millimetre telescope to search for emission from the \transe\/
  (85.5-GHz) and \transa\/ (86.6-GHz) transitions of methanol.  The
  search was targeted towards 22 star formation regions which exhibit
  maser emission in the 107.0-GHz $3_{1}\mbox{-}4_{0}\mbox{~A}^{+}$
  methanol transition, as well as in the 6.6-GHz
  $5_{1}\mbox{-}6_{0}\mbox{~A}^{+}$ transition characteristic of
  class~II methanol maser sources.  A total of 22 regions were
  searched at 85.5~GHz resulting in 5 detections, of which 1 appears
  to be a newly discovered maser.  For the 86.6-GHz transition
  observations were made of 18 regions which yielded 2 detections, but
  no new maser sources.  This search demonstrates that emission from
  the \transe\/ and \transa\/ transitions is rare.  Detection of
  maser emission from either of these transitions therefore indicates
  the presence of special conditions, different from those in the
  majority of methanol maser sources.  We have observed temporal
  variability in the 86.6-GHz emission towards 345.010+1.792, which
  along with the very narrow line width, confirms that the emission is
  a maser in this source.
  
  We have combined our current observations with published data for
  the 6.6-, 12.1-, 85.5-, 86.6-, 107.0-, 108.8- and 156.6-GHz
  transitions for comparison with the maser model of Sobolev \&
  Deguchi (1994).  Both detections and nondetections are useful for
  setting limits on the physical conditions in star forming regions
  which contain methanol maser emission.  This has allowed us to
  estimate the likely ranges of dust temperature, gas density, and
  methanol column density, both for typical methanol maser sources and
  for those sources which also show 107.0-GHz emission.  The gas
  temperature can also be estimated for those sources exhibiting
  masers at 85.5 and/or 86.6~GHz.

\end{abstract}

\begin{keywords}
masers -- stars:formation -- ISM: molecules -- radio lines : ISM
\end{keywords}

\section{Introduction}

More than 30 different transitions of methanol have been observed to
exhibit interstellar maser emission.  The maser transitions are empirically
classified into two groups (named class~I and class~II) on the basis
of their association with high-mass star formation regions \cite{M91a}.
Class II transitions are closely associated with signposts of
high-mass star formation such as ultra-compact \ionhy\/ regions, OH
masers and strong far infra-red emission.  Class I masers occur offset
from star formation regions and are believed to be collisionally
pumped at the interface between outflows and the parent molecular
cloud \cite{M91a}.

In contrast to OH and \water\/ masers, class~II methanol maser
emission is only found associated with high-mass star formation,
making it a useful signpost of such regions.  In particular the
6.6-GHz $5_{1}\mbox{-}6_{0}\mbox{~A}^{+}$ transition has been detected
towards more than 550 sites throughout our Galaxy
\cite{M91b,MGN92,CVEWN95,EVMNDP96,WHRB97,SVKVPTC99,SKHKP02}.  More
than 120 of the regions which exhibit 6.6-GHz maser emission also
exhibit maser emission in the 12.1-GHz $2_{0}\mbox{-}3_{-1}\mbox{~E}$
transition \cite{CVEN95}.  Approximately 175 known 6.6-GHz maser
sources have been searched in the $3_{1}\mbox{-}4_{0}\mbox{~A}^{+}$
transition at 107.0~GHz, with maser emission detected towards 25
regions \cite{VDKSBW95,VESKOV99,CYBC00,MB02}.  In contrast, the scale
of searches and the number of detections for the other 20 or so
class~II masing transitions have been very limited.

Sobolev \& Deguchi \shortcite{SD94} developed the first theoretical
model to successfully reproduce the high brightness temperatures that
characterise strong class~II methanol masers (referred to hereafter as
the SD model).  In the SD model the masers are pumped by infra-red
radiation from warm dust close to the star forming region, where the
gas phase abundance of methanol is greatly enriched following grain
mantle evaporation.  The SD model finds that the 6.6- and 12.1-GHz
transitions produce strong maser emission over a wide range of
physical conditions, providing an explanation as to why maser emission
is common in these two transitions \cite{SCG97a}.  It also predicts
weaker maser emission from a variety of methanol transitions
\cite{SCG97b}, including all the transitions discovered previously
\cite{WWSJ84,WWMH85,HBM89,WHDH93,SKV95} and subsequently
\cite{VESKOV99,CSECGSD01}.

There is considerable evidence that many of the class~II methanol
maser transitions are spatially coincident, arising from the same
region.  The 6.6- and 12.1-GHz methanol masers have been shown to be
coincident at the milliarcsecond level towards a number of sources
\cite{MRPMW92,NWCWG93,MBC00}.  The higher frequency weak maser
transitions typically align in velocity with 6.6- and 12.1-GHz
spectral features, but definitively proving that they are coincident
is more difficult.  However, recent BIMA observations of the 86.6- and
86.9-GHz masers in W3(OH) show that they are coincident with the
strongest 107.0 GHz masers to within the 0\farcs1 accuracy of the
observations \cite{SSECMOG01}.  These observations support the
proposition that the class~II methanol maser transitions are excited
simultaneously, which if true means that their observation provides tight
constraints for maser pumping models.

Many transitions become masers simultaneously in the SD model, so that
when masers at several frequencies coincide, the best fitting model
conditions can be estimated via multi-transition analysis.  This
provides a probe of the physical conditions in the maser region.  The
maser transitions which are strongly inverted only over a small range
of model conditions are potentially much more useful for setting
constraints on physical conditions than are the stronger, more
ubiquitous maser transitions.  For example in the modeling of Sobolev
\etal\/ the 85.5-GHz transition is predicted to be strongest for gas
temperatures less than 50K and gas densities in the range $10^6$ --
$10^8$~cm$^{-3}$, while the 86.6-GHz transition favours higher
temperatures and lower densities.  The observations of W3(OH) by
Sutton \etal\/ \shortcite{SSECMOG01} detected maser emission from the
same location in the 86.6-, 86.9- and 107.0-GHz transitions, but only
thermal or quasi-thermal emission from the 85.5-GHz transition,
consistent with a warm gas model.  In contrast, towards the southern
star forming region 345.010+1.792, 85.5- 86.6-, 86.9-, 107.0- and
108.1-GHz masers have all been detected from the same velocity range
\cite{VESKOV99,CSECGSD01}.

Previous searches for the 85.5- and 86.6-GHz methanol maser
transitions have been made towards 30 and 35 sources respectively
\cite{CSECGSD01,SSECMOG01,MB02}.  In this work we report the results
of observations of a further 15 and 10 sources not previously searched
for 85.5- and 86.6-GHz methanol maser emission respectively.  Both
weak maser and quasi-thermal emission have been detected, and it is
not always easy to distinguish between the two in single dish
observations.  The 85.5-GHz emission is thought to be of maser origin
in 345.010+1.792, 9.621+0.196, 29.95-0.02 and DR21(OH), while the
86.6-GHz emission shows maser characteristics towards W3(OH),
345.010+1.792 and W51-IRS1.  The search described here forms part of a
program to observe a number of the rarer methanol maser transitions in
a consistent sample of sources, selected on the basis of their
107.0-GHz maser emission. Of the 25 known 107.0-GHz masers, 22 can be
observed with the Mopra 22-m telescope, and we have searched for 85.5-
and 86.6-GHz methanol masers towards those 107.0-GHz masers not
targeted in previous surveys.  Since only 4 of the 7 previous
detections were associated with 107.0-GHz emission, this strategy will
provide useful information for modelling the target sources, but does
not encompass all candidates for new 85.5- and 86.6-GHz emission.

\section{Observations and Data reduction}

The observations were made between 2002 July 31 and August 13 using
the Australia Telescope National Facility (ATNF) 22-m millimetre
antenna at Mopra.  At 86~GHz the FWHM antenna beamwidth was
39\arcsec\/.  The sources were observed in position switching mode,
with 229 seconds spent at the onsource position and 229 seconds offset
by -1\arcmin\/ in declination.  For the majority of sources this
procedure was repeated 10 times giving a total onsource integration
time of 38~minutes, which typically yielded an RMS of 0.7~Jy.  Two
different transitions were observed, the \transe\/ and \transa\/
transitions for which the adopted rest frequencies were 85.568084 GHz
and 86.615578 GHz respectively \cite{TUTT95}.  The data were collected
using a 2-bit digital autocorrelation spectrometer configured with
1024 channels spanning a 64-MHz bandwidth.  For an observing frequency
of 86~GHz this configuration yields a natural weighting velocity
resolution of 0.26~\kms , or 0.44~\kms\/ after Hanning smoothing.

The antenna pointing was checked approximately every two hours through
observations of nearby strong SiO masers and showed an RMS scatter of
the order of 10\arcsec .  The system temperature was determined by
inserting an ambient temperature load which was assumed to have a
temperature of 295K.  The measured values varied between 210 and 300K
during the observations depending upon the weather conditions and
elevation.  This method of calibration also corrects for atmospheric
absorption \cite{KU81} and taking into account pointing inaccuracies
the absolute flux density scale of the observations is believed to be
accurate to 20 per cent.  The amplitude scale was converted from
antenna temperature to Jy assuming a sensitivity of 30~JyK$^{-1}$.  The
data were processed using the {\sc spc} reduction package.  Quotient
spectra were formed for each on/off pair of observations, which were
then averaged together, a polynomial baseline fitted and subtracted
and the velocity and amplitude scale calibrated.  

An erroneous observing procedure meant that at each of 85.5- and
86.6-GHz three sources were observed with the telescope offset by
+20\arcsec\/ in right ascension (approximately half the FWHM of the
antenna beam at this frequency) from the nominal position.  The upper
limit listed in Tables~\ref{tab:meth85} \& \ref{tab:meth86} for the
sources observed at an offset position has been set to 9 ($3\times3)$
times the RMS noise level in the final spectrum as a conservative
estimate of the detection limit.

\section{Results}

\subsection{\transe\/ (85.5~GHz)}
\begin{figure*}
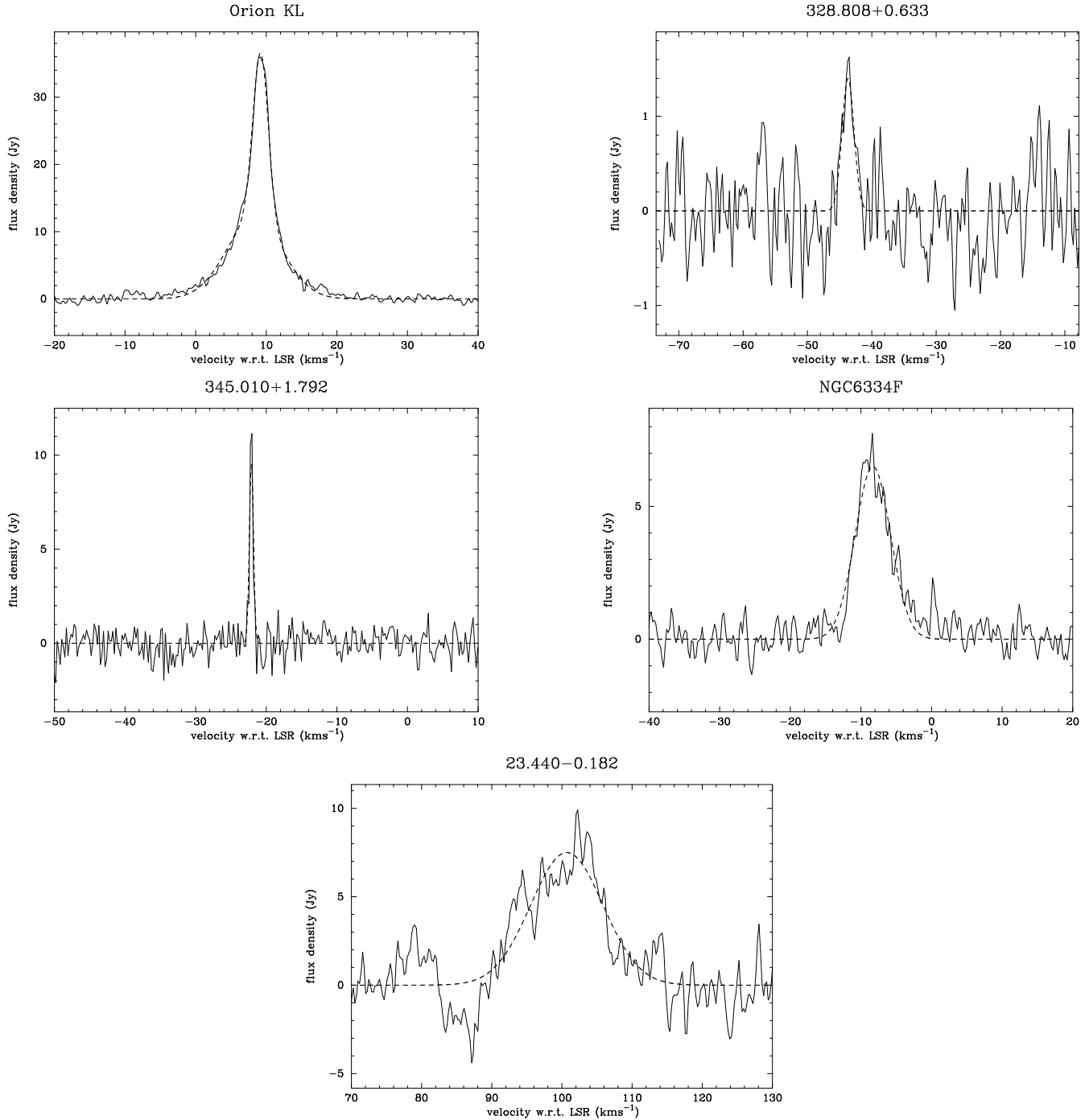

  \specdfig{orion_kl_85_5}{328.808_85_5}
  \specdfig{345.010_85_5}{351.417_85_5}
  \specsfig{23.440_85_5}
\caption{Spectra of the sources detected at 85.5 GHz.  With the exception of
345.010+1.792 all spectra have been Hanning smoothed.}
\label{fig:meth85}
\end{figure*}

A total of 22 sources were observed at 85.5 GHz and these are listed
in Table~\ref{tab:meth85}.  With the exception of Orion~KL each of
these sources shows maser emission in the 6.6- and 107.0-GHz
transitions.  The observations of Orion~KL were used to test the
system performance and to aid comparison of the calibration with the
results of Cragg \etal\/ \shortcite{CSECGSD01} and Minier \& Booth
\shortcite{MB02}.  Emission from the 85.5-GHz \transe\/ transition of
methanol was detected towards 5 sources.  The spectra of these sources
are shown in Fig.~\ref{fig:meth85}.  Three of the line profiles are
broad (Orion~KL, NGC6334F and 23.440-0.182) and so the emission is
thermal, or quasi-thermal in origin.  The emission from 345.010+1.792
is very strong and narrow, and as it aligns with the maser emission at
other frequencies and is offset from the velocity range of the thermal
emission in this source, it is clearly also masing in this transition.
The 85.5-GHz emission towards 328.808+0.633 is narrow, but much weaker
than 345.010+1.792 and so from these observations we cannot be certain
that it is a maser in this transition.  However, the velocity of the
emission coincidences exactly with the velocity of the 107-GHz
methanol maser in this source, adding support to the argument that
328.808+0.633 is an 85.5-GHz maser.  The observation of 9.621+0.196
with 3-$\sigma$=1.5~Jy was not sufficiently sensitive to confirm the
previous detection of this source with a peak flux density of 1.2~Jy
\cite{CSECGSD01}.  For modelling we have used the flux density
measured by Cragg \etal\/ rather than the current observations.

\begin{table*}
  \caption{A list of sources observed at 85.5-GHz.  For sources where 
  emission was detected the values listed are those of Gaussian profiles
  fitted to the spectra.  For sources where no emission was detected the 
  value listed is 3 x the RMS noise level in the Hanning smoothed spectrum.  
  Upper limits indicated with a $^\dagger$ were observed offset by 20\arcsec\/
  in right ascension and have been multiplied by a further factor of 3 to
  account for reduced sensitivity.}
  \begin{tabular}{lllrccc} \hline
    {\bf Name} & {\bf R.A.(J2000)} & {\bf Dec.(J2000)}                      & 
    {\bf Integration} & {\bf Flux Density} & {\bf Velocity} & {\bf FWHM}    
    \\
               & {\bf (h~m~s)}     & {\bf (\degrees\/ \arcmin\/ \arcsec\/)} & 
    {\bf Time (min)}  & {\bf (Jy)}         & {\bf (\kms\/)} & {\bf (\kms\/)}
    \\ [2mm] \hline
    Orion KL       & 05:35:14.5 & --05:22:30 &   57 & 25.0   & 9.2   & 2.6 \\
                   &            &            &      & 11.8   & 8.3   & 9.4 \\
    188.946+0.886  & 06:08:53.3 & +21:38:29  &   38 & $<$2.1 &       &     \\
    192.600--0.048 & 06:12:54.0 & +17:59:24  &   57 & $<$1.8 &       &     \\
    310.144+0.760  & 13:51:58.5 & --61:15:42 &   76 & $<$1.4 &       &     \\
    318.948--0.196 & 15:00:55.4 & --58:58:53 &   38 & $<$2.0 &       &     \\
    323.740--0.263 & 15:31:45.5 & --56:30:50 &   38 & $<$1.8 &       &     \\
    327.120+0.511  & 15:47:32.7 & --53:52:38 &   76 & $<$1.5 &       &     \\
    328.808+0.633  & 15:55:48.5 & --52:43:07 &   76 & 1.4    & -43.7 & 2.2 \\
    336.018--0.827 & 16:35:09.3 & --48:46:47 &   76 & $<$1.3 &       &     \\
    339.884--1.259 & 16:52:04.7 & --46:08:34 &  114 & $<$1.1 &       &     \\
    340.785--0.096 & 16:50:14.8 & --44:42:26 &   38 & $<$1.8 &       &     \\
    345.003--0.223 & 17:05:10.9 & --41:29:06 &   38 & $<$2.0 &       &     \\
    345.010+1.792  & 16:56:47.6 & --40:14:26 &   37 & 9.5    & -22.1 & 0.7 \\ 
    345.504+0.348  & 17:04:22.9 & --40:44:22 &   38 & $<4.7^\dagger$ &     & \\
    348.703--1.043 & 17:20:04.1 & --38:58:31 &   38 & $<$1.5 &       &     \\
    NGC6334F       & 17:20:53.4 & --35:47:01 &   37 & 6.5    & -8.3  & 5.9 \\
    353.410--0.360 & 17:30:26.2 & --34:41:46 &   38 & $<$2.1 &       &     \\
    9.621+0.196    & 18:06:14.7 & --20:31:32 &   38 & $<$1.5 &       &     \\
    12.909--0.260  & 18:14:39.5 & --17:52:00 &   34 & $<5.0^\dagger$ &     & \\
    23.010--0.411  & 18 34 40.3 & --09:00:38 &   38 & $<5.1^\dagger$ &     & \\
    23.440--0.182  & 18:34:39.2 & --08:31:24 &   19 & 7.5    & 100.7 & 12.1 \\
    35.201--1.736  & 19:01:45.5 & +01:13:35  &   91 & $<$1.1 &       &    \\ \hline
  \end{tabular}
  \label{tab:meth85}
\end{table*}

Gaussian profiles have been fitted to each of the sources where emission
was detected and parameters are listed in Table~\ref{tab:meth85}.

\subsection{\transa\/ (86.6~GHz)}

\begin{figure*}
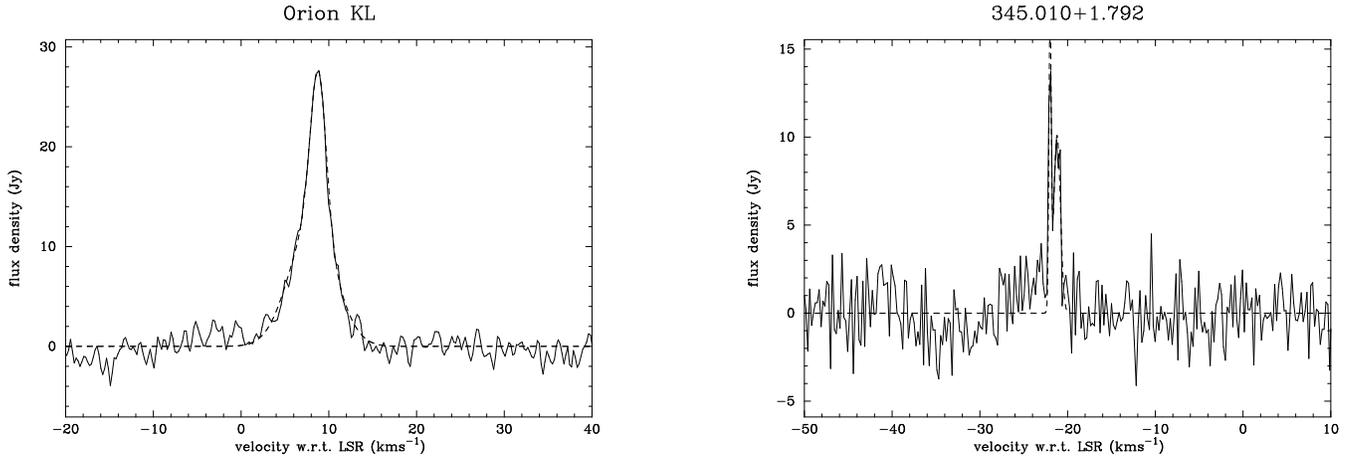

  \specdfig{orion_kl_86_6}{345.010_86_6}
\caption{Spectra of the sources detected at 86.6 GHz}
\label{fig:meth86}
\end{figure*}

The 86.6-GHz \transa\/ transition was detected towards 3 of the 18
sites observed (see Table~\ref{tab:meth86}).  The spectra of sources
where emission was detected (with the exception of NGC6334F) are shown
in Fig~\ref{fig:meth86}.  Maser emission was only detected in one
source, 345.010+1.792, with thermal emission detected towards
Orion~KL and NGC6334F.  The observation of NGC6334F was at an offset
position and is of poor quality compared to previously published
spectra \cite{CSECGSD01}, so we have not included it in
Fig~\ref{fig:meth86}.  Marginal narrow emission was detected near the
velocity of the 107.0-GHz maser peak towards two other sources
(323.740-0.263 \& 339.884-1.259) each with a significance of
approximately 4-$\sigma$.

\begin{table*}
  \caption{A list of sources observed at 86.6-GHz.  For sources where 
  emission was detected the flux density and velocity are the values at
  the peak.  For sources where no emission was detected the value listed
  is 3 x the RMS noise level in the Hanning smoothed spectrum.  Upper
  limits indicated with a $^\dagger$ were observed offset by 20\arcsec\/
  in right ascension and have been multiplied by a further factor of 3 to
  account for reduced sensitivity.}
  \begin{tabular}{lllrccc} \hline
    {\bf Name} & {\bf R.A.(J2000)} & {\bf Dec.(J2000)}                      & 
    {\bf Integration} & {\bf Flux Density} & {\bf Velocity} & {\bf FWHM}
    \\
               & {\bf (h~m~s)}     & {\bf (\degrees\/ \arcmin\/ \arcsec\/)} & 
    {\bf Time (min)}  & {\bf (Jy)}         & {\bf (\kms\/)} & {\bf (\kms\/)}
    \\ [2mm] \hline
    Orion KL       & 05:35:14.5 & --05:22:30 &   19 & 15.8   & 8.8   & 2.1 \\
                   &            &            &      & 12.3   & 8.1   & 6.1 \\
    188.946+0.886  & 06:08:53.3 & +21:38:29  &   34 & $<$2.4 &       & \\
    192.600--0.048 & 06:12:54.0 & +17:59:24  &   34 & $<$2.2 &       & \\ 
    310.144+0.760  & 13:51:58.5 & --61:15:42 &   38 & $<$2.1 &       & \\
    318.948--0.196 & 15:00:55.4 & --58:58:53 &   38 & $<$2.2 &       & \\
    323.740--0.263 & 15:31:45.5 & --56:30:50 &   76 & $<$1.9 &       & \\ 
    327.120+0.511  & 15:47:32.7 & --53:52:38 &   38 & $<$2.2 &       & \\
    339.884--1.259 & 16:52:04.7 & --46:08:34 &   34 & $<$2.4 &       & \\ 
    340.054--0.244 & 16:48:13.9 & --45:21:44 &   34 & $<$2.3 &       & \\
    340.785--0.096 & 16:50:14.8 & --44:42:26 &   34 & $<$1.9 &       & \\
    345.010+1.792  & 16:56:47.6 & --40:14:26 &   28 & 16.4   & -22.0 & 0.3 \\
                   &            &            &      & 10.1   & -21.2 & 0.9 \\
    345.504+0.348  & 17:04:22.9 & --40:44:22 &   31 & $<$2.4 &       & \\
    348.703--1.043 & 17:20:04.1 & --38:58:31 &   38 & $<$2.7 &       & \\
    NGC6334F       & 17:20:53.4 & --35:47:01 &   38 & 4.0$^\dagger$ &       & \\ 
    353.410--0.360 & 17:30:26.2 & --34:41:46 &   38 & $<$2.0 &       & \\
    9.621+0.196    & 18:06:14.7 & --20:31:32 &   38 & $<6.6^\dagger$ &       & \\
    12.909--0.260  & 18:14:39.5 & --17:52:00 &   38 & $<8.6^\dagger$ &       & \\
    23.440--0.182  & 18:34:39.2 & --08:31:24 &   38 & $<$4.1 &       & \\ \hline
  \end{tabular}
  \label{tab:meth86}
\end{table*}

\section{Discussion}

As well as assessing the prevalence of masers in the 85.5-GHz
\transe\/ and 86.6-GHz \transa\/ transitions of methanol, this work
sets constraints on maser modelling.  According to models of maser
pumping, the detection of methanol maser emission in a star formation
region signals the presence of special physical conditions, probably
associated with a particular evolutionary phase.  The common 6.6- and
12.1-GHz methanol masers are excited for a wide range of model
parameters and so do not tightly constrain the physical conditions
within the region.  In contrast, modelling suggests that many of the
other, rarer methanol maser transitions are inverted over a more
restricted range of parameters.  The 85.5- and 86.6-GHz transitions
each predominates over a different range of physical conditions
\cite{CSECGSD01}, making them potentially useful probes of gas
temperature and density.

Table~\ref{tab:fluxes} summarises the flux densities in 7 methanol
maser transitions of the 25 star forming regions currently known to
harbour 107.0-GHz maser emission \cite{CYBC00,VDKSBW95}.  For the
6.6-, 12.1-, 107.0-, 108.8- and 156.6-GHz transitions the flux
densities cited are at the velocity of the 107.0-GHz peak and have
been determined by eye from published spectra.  For many of the
sources detected at 108.8 and 156.6~GHz, thermal emission with no
apparent maser component covers the velocity range of the 107.0-GHz
peak.  Where this occurs the upper limit listed in the table is marked
with an asterisk and has been taken to be half the amplitude of the
thermal emission at that velocity.  This approach has also been used
in cases where thermal emission was detected from other transitions.
The most noticeable feature of Table~\ref{tab:fluxes} is that the
85.5-, 86.6-, 108.8- and 156.6-GHz transitions are all uncommon (in
the presence of 6.6-, 12.1- and 107.0-GHz masers).  By investigating
the range of physical conditions favouring each of these weaker
transitions, we are able to distinguish between common and uncommon
conditions within high-mass star formation regions.



\begin{table*}
  \caption{The flux density at the velocity of the peak in the 107.0-GHz 
    transition of the 6.6-, 12.1-, 85.5-, 86.6-, 107.0-, 108.8- \& 156.6-GHz 
    methanol transitions towards all star formation regions with known 
    107.0-GHz methanol masers.  The 85.5- and 86.6-GHz data is taken from this
    work, except where indicated by a $^\dagger$.  The information for the
    other transitions has been taken from the literature.  Where upper limits
    are quoted they are 3 times the RMS noise level in the spectra.  
    Transitions where thermal emission is detected are indicated with a $^*$ 
    and the upper limited listed is 50\% of the flux density of the thermal 
    emission at the 107.0-GHz peak velocity.  References : 
    1=Batrla \etal\/ \protect\shortcite{BMMW87};
    2=Caswell \etal\/ \protect\shortcite{CVEWN95};
    3=Caswell \etal\/ \protect\shortcite{CVEN95};
    4=Caswell \etal\/ \protect\shortcite{CYBC00};
    5=Cragg \etal\/ \protect\shortcite{CSECGSD01};
    6=Koo \etal\/ \protect\shortcite{KWHB88};
    7=Mehringer, Zhou \& Dickel \protect\shortcite{MZD97};
    8=Menten \protect\shortcite{M91b};
    9=Minier \& Booth \protect\shortcite{MB02};
    10=Slysh, Kalenskii \& Val'tts \protect\shortcite{SKV95};
    11=Sutton \etal\/ \protect\shortcite{SSECMOG01};
    12=Val'tts \etal\/ \protect\shortcite{VDKSBW95};
    13=Val'tts \etal\/ \protect\shortcite{VESKOV99};
    14=Walsh \etal\/ \protect\shortcite{WHRB97}
  }
  \begin{tabular}{lrrrrrrrrl}
                 &                & \multicolumn{7}{c}{{\bf Flux Density}}
                                              &                  \\
    {\bf Source} & {\bf Velocity} & {\bf 6.6-GHz}      & {\bf 12.1-GHz}     & 
      {\bf 85.5-GHz}     & {\bf 86.6-GHz}     & {\bf 107.0-GHz}    & 
      {\bf 108.8-GHz}    & {\bf 156.6-GHz}    & {\bf references} \\
                 & {\bf (\kms)}   & {\bf (Jy)}         & {\bf (Jy)}         & 
      {\bf (Jy)}         & {\bf (Jy)}         & {\bf (Jy)}         & 
      {\bf (Jy)}         & {\bf (Jy)}         &                  \\ \hline
  W3(OH)          &  -43.3 & 3000 & 600    &  $<0.7^\dagger$  & $6.7^\dagger$     & 72   &
  $<$0.6     & $<$9     & 8,10,11 \\
  188.946+0.886   &   10.9 & 495  & 235    &  $<$2.1  & $<$0.8  & 15.5 & 
  $<$4.8     & $<$2     & 2,3,4,5,13 \\
  192.600-0.048   & 4.2    & 72   & $<$0.4 &  $<$1.8  & $<$2.2  & 5.8  &
  $<$4.5     & $<$3     & 2,3,4,13 \\
  310.144+0.760   & -56    & 130  & 114    &  $<$1.4  & $<$2.1  & 23   &
             & $<$3     & 4 \\ 
  318.948--0.196  & -34.2  & 780  & 180    &  $<$2.0  & $<$2.2  & 5.7  &
  $<3^{*}$   & 2.4      & 2,3,4,13 \\
  323.740--0.263  & -50    & 2000 & 500    &  $<$1.8  & $<$1.9  & 12.5 &
  $<$5.4     & $<$4$^*$ & 2,3,4,13 \\ 
  327.120+0.511   & -89.8  & 25   & 5      &  $<$1.5  & $<$2.2  & 9.2  &
             & $<$2.8   & 2,3,4 \\
  328.808+0.633   & -43.5  & 380  & 7      &  1.6     & $<2.4^\dagger$ & 5.5  &
  $<6^{*}$   & $<$8$^*$ & 2,3,4,5,13 \\ 
  336.018--0.827  & -40    & 40   & 25     &  $<$1.3  & $<2.7^\dagger$ & 6    &
  $<$5.7     & $<$5$^*$ & 4,5,13,14 \\ 
  339.884--1.259  & -39.0  & 1820 & 850    &  $<$1.1  & $<1.7^\dagger$ & 90   &
  $<$5.4     & 6        & 2,3,4,13 \\
  340.054--0.244  & -62.8  & 0.3  & $<$1   &          & $<$2.3  & 2.9  &
             & $<$2     & 2,3,4 \\
  340.785--0.096  & -105.9 & 144  & 43     & $<$1.8   & $<$1.9  & 6.1  &
             & $<$2     & 2,3,4 \\
  345.003--0.223  & -26.9  & 160  & $<$1   &  $<$2.0  & $<$2.9  & 3.5  &
  $<1.5^{*}$ & $<$6$^*$ & 2,3,4,5,13 \\ 
  345.010+1.792   & -21.7  & 330  & 300    &  11.2    & 13.8    & 82   &
  7.5        & 18       & 2,3,4,13 \\
  345.504+0.348   & -17.7  & 174  & 4.7    &  $<$4.7  & $<$2.4  & 2.3  &
  $<2^{*}$   & $<$4$^*$ & 2,3 \\ 
  348.703--1.043  & -3.3   & 60   & 34.5   &  $<$1.5  & $<$2.7  & 7.6  &
             & $<$1.5   & 2,3,4 \\
  NGC6334F        & -10.2  & 3300 & 1000   & $<2.5^*$ & $<1.0^{*\dagger}$ & 14.8 &
             & $<$17$^*$ & 2,3,4,5 \\ 
  353.410--0.360  & -20    & 90   & 5      &  $<$2.1  & $<$2.0  & 5.5  &
             & $<$2     & 2,3,4 \\
  9.621+0.196     & -0.5   & 100  & 10     &  $<$1.5  & $<2.3^\dagger$ & 22   &
  $<3^{*}$   & $<$3$^*$ & 2,3,4,5,13 \\ 
  12.909--0.260   & 39.5   & 317  & 11.5   &  $<$3.8  & $<8.6^\dagger$ & 5.5  &
  $<3^{*}$   & $<$3$^*$ & 2,3,4,5,13 \\ 
  23.010--0.411   & 75.9   & 405  & 28     &  $<3.8^\dagger$ &         & 5.2  &
  $<$5.1     & $<$2     & 2,3,4,5,13 \\ 
  23.440--0.182   & 97.2   & 23   & 9      & $<4.0^*$ & $<$4.1  & 4.4  &
             & $<$2     & 2,3,4 \\
  35.201--1.736   & 42     & 560  & 25     &  $<$1.1  & $<2.3^\dagger$ & 24   &
  $<$4.8     & 4.6      & 2,3,4,5 \\ 
  Cep A           & -2.2   & 1420 & $<$7   &  $<5.0^\dagger$ & $<5.0^\dagger$ & 16   &
  $<$5.0     & $<$3     & 6,7,8,9,10 \\ 
  NGC7538         & -56.3  & 346  & 200    &  $<5.0^\dagger$ & $<5.0^\dagger$ & 17   &
  $<4.0^{*}$ &          & 1,8,9,12 \\ \hline
  \end{tabular}
\label{tab:fluxes}
\end{table*}

\subsection{Variability in 345.010+1.792}

\begin{figure}
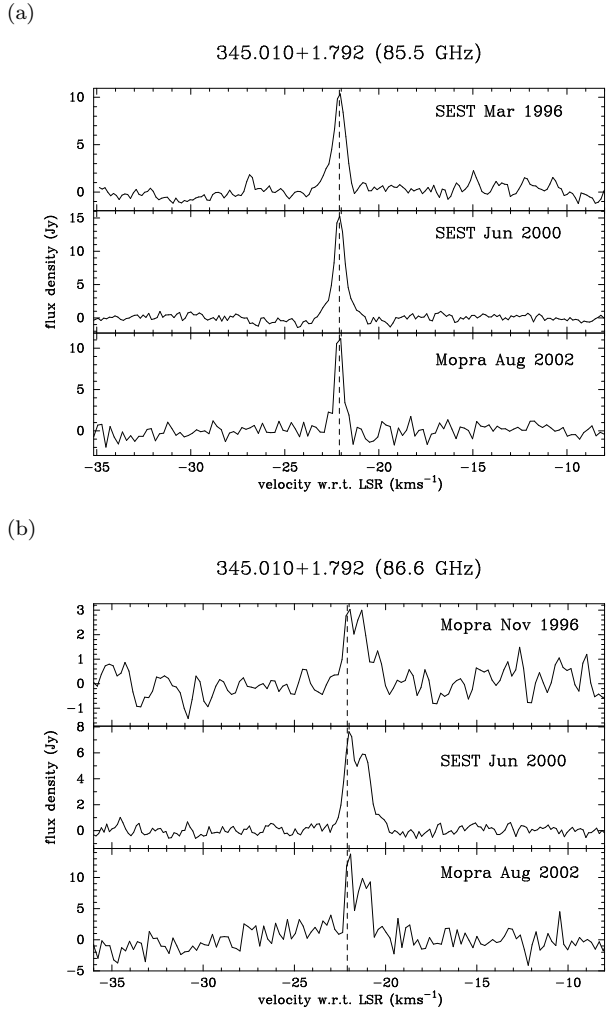

  \speclfig{(a)}{345.010_stack_85}
  \speclfig{(b)}{345.010_stack_86}
  \caption{(a) Spectra of the 85.5-GHz maser emission in 345.010+1.792 at 3 
    different epochs.  The velocity of the spectrum taken at SEST in
    2000 has been shifted by -0.3 \kms\/ and that of the spectrum from
    Mopra in 2002 has been shifted by -1.3 \kms. (b) Spectra of the
    86.6-GHz maser emission in 345.010+1.792 at 3 different epochs.
    The velocity of the spectrum taken at Mopra in 1996 has been
    shifted by 0.3 \kms\/ and that of the spectrum from SEST in 2000
    has been shifted by -0.3 \kms .  The dotted line at -22.1~\kms\/
    shows the velocity of the peak of the 85.5-GHz emission.}
\label{fig:345_stack}
\end{figure}

The 85.5- and 86.6-GHz methanol masers in 345.010+1.792 were
discovered by Cragg \etal\/ \shortcite{CSECGSD01}.  Our survey
confirms the exceptional nature of this source, which remains the
prime example of masers at these frequencies.  Comparison of spectra
for the same transitions obtained at different epochs or with
different telescopes allows us to examine variability in these
transitions, although this is complicated by uncertainties in the
velocity and flux density calibration.  Figure~\ref{fig:345_stack}a
shows the 85.5-GHz emission in 345.010+1.792 at 3 different epochs
covering a period of 6 years.  The single peaked line profile appears
to have narrowed slightly over the period, but it is not possible to
quantify the variability as some is likely to be due to calibration
uncertainties.  Figure~\ref{fig:345_stack}b shows the 86.6-GHz
emission in 345.010+1.792 at 3 epochs covering essentially the same
time range as the 85.5-GHz observations.  The 86.6-GHz transition
shows a clear double peaked profile and comparison of the relative
intensities of these two features shows significant changes over a 6
year interval.  The amplitude calibration of the 1996 observations
from Mopra was uncertain by as much as a factor of 2 \cite{CSECGSD01}
and comparison of the 86.6-GHz spectrum of Orion KL from 1996 with
that from the current observations shows the 1996 peak flux density to
be lower by a factor of 1.8.  The emission from Orion is thought to be
quasi-thermal in origin and should therefore not vary, which suggests
that the flux densities from the 1996 Mopra 86.6-GHz observations
should be scaled in amplitude by a factor of 1.8.  Note that all the
spectra in Fig.~\ref{fig:345_stack} have been aligned with the peak
emission velocity in observations made in October 2002 using the newly
upgraded Australia Telescope Compact Array (ATCA) at 3mm (Minier,
Burton \& Wong, in prep).  This required small, but significant shifts
in some of the spectra, as detailed in the Figure caption, which can
be attributed to an error in the AOS frequency calibration for the
SEST 2000 observations, and intermittent errors in the velocity scale
which are known to effect Mopra observations.

The very narrow line width, offset velocity from the thermal emission
in other transitions towards this source and clear temporal
variability provides conclusive evidence that the 86.6-GHz emission in
345.010+1.792 is a maser.  It should also be noted that the strongest
emission from the 85.5- and 86.6-GHz transitions are coincident in
velocity to within 0.1~\kms\/.  The recent imaging of the 85.5- and
86.9-GHz methanol masers in 345.010+1.792 by Minier \etal\/ has shown
them to be spatially coincident at the arcsecond level at the same
position ($\alpha_{J2000}$=16:56:47.56, $\delta_{J2000}$=-40:14:26.3)
as the 6.6-, 12.1- and 107.0 GHz masers.

\subsection{Modelling} \label{sec:modelling}

The observations reported here, combined with the search of Minier \&
Booth \shortcite{MB02}, demonstrate that class~II methanol masers at
85.5 and 86.6~GHz are weak ($\le 14$~Jy), and rarely detected at the
sensitivity of the current experiment.  We can use the results of
maser modelling calculations to distinguish between conditions in those
few sources which display masers at these frequencies, the remaining
target sources with methanol masers at both 6.6 and 107.0~GHz, and the
majority of 6.6-GHz methanol maser sources which lack millimetre
wavelength maser counterparts.

We have used the model of Sobolev \& Deguchi \shortcite{SD94} to map
out the prevalence of the various class~II methanol masers as a
function of the model parameters representing physical conditions.
Calculations involving this model were previously reported by Sobolev
\etal\/ \shortcite{SCG97a,SCG97b}, Cragg \etal\/
\shortcite{CSECGSD01}, Sutton \etal\/ \shortcite{SSECMOG01}, and
Cragg, Sobolev \& Godfrey \shortcite{CSG02}.  Altogether 1131 energy
levels of methanol (with torsional states $\vt=0,1,2$ for both A and E
symmetry species) were included \cite{MGH99}.  The maser pumping is
governed by infrared radiation from warm dust at temperature \Td,
which promotes the methanol molecules to their torsionally excited
states.  We assumed a dust filling factor of 0.5 and opacity
$(\nu/10^{13})^2$, where $\nu$ is the frequency in Hz.  This is
consistent with the observations of De Buizer, Pi\~na \& Telesco
\shortcite{DPT00} that the optical depth at 9.7~$\mu$m of mid-infrared
emission associated with methanol masers can be as large as 1.
Radiative transfer was treated in the large velocity gradient (LVG)
approximation, augmented by a beaming factor $\beam=10$, defined as
the ratio of optical depths parallel and perpendicular to the line of
sight.  This represents the assumed elongation of the maser region
towards the observer.  In this treatment the maser amplification and
hence the brightness is governed by \beam\scdm, representing the
column density of methanol along the line of sight divided by the line
width of the emission.  In terms of observable parameters, multiplying
\scdm\ by the typical line width of a methanol maser (0.5~\kms)
and the beaming factor gives the methanol column density.  The
parameter \scdm\ is referred to as the specific column density of
methanol.  The masers develop in gas of temperature \Tk\ and density
\nH, where the methanol molecules undergo collisions with molecular
hydrogen.  Collisional excitation rates are based on the model of Peng \& Whiteoak \shortcite{PW93}, which uses propensity rules derived from a small number of laboratory measurements.  The masers
amplify radiation from a background uc\HII\ region, here assumed to
have brightness temperature $10^4(1-\rm{exp}[-(1.2\times
10^{10}/\nu)^2])$~K, and geometrical dilution 0.002.  The model produces methanol masers in the absence of a background uc\HII\ region, but with lower brightness temperature in the centimetre wavelength masers than when the uc\HII\ region radiation is included.

\begin{figure*}
  \psfig{file=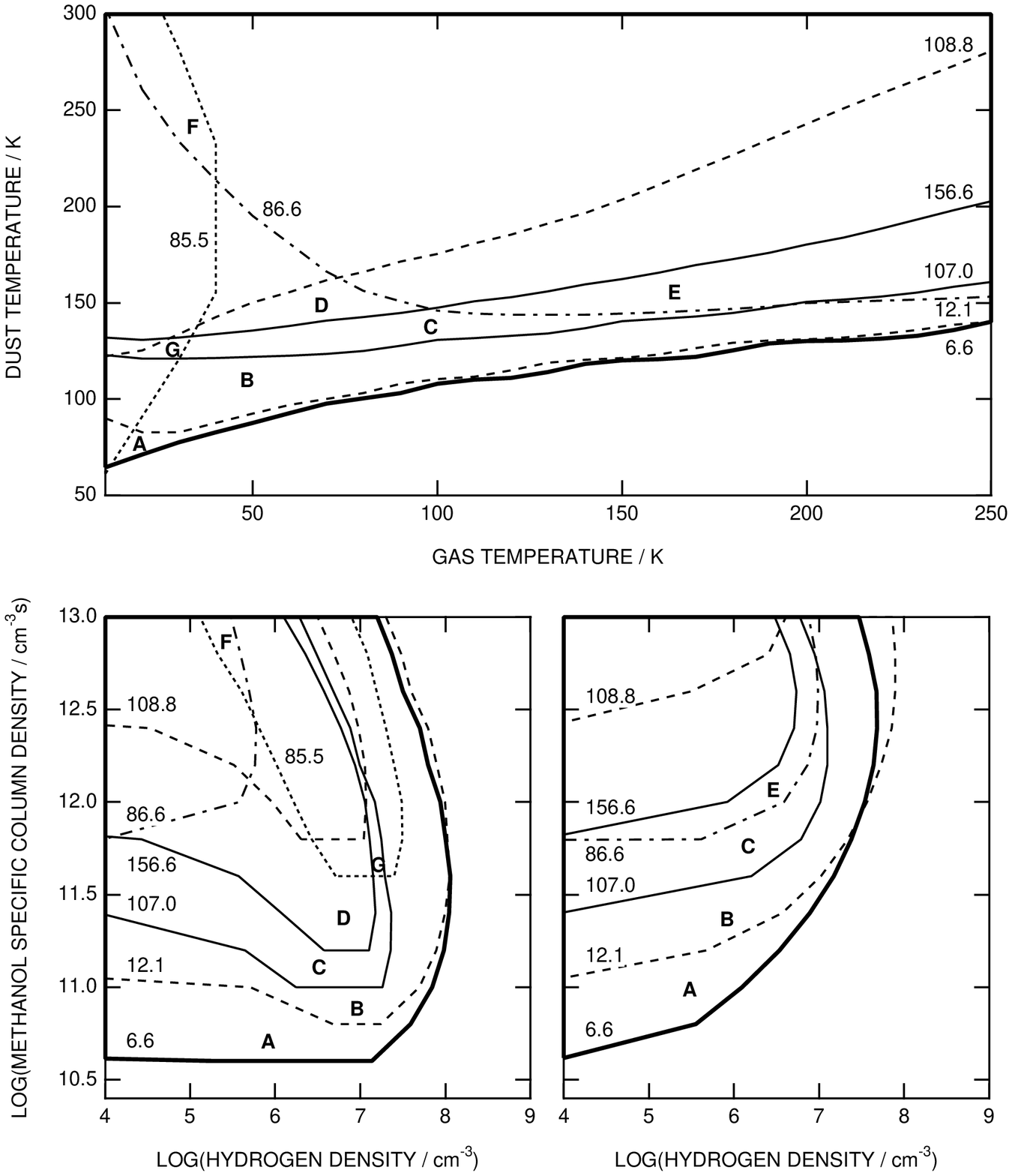,height=0.82\textwidth}
  \caption{Contour diagrams showing regions where selected class~II methanol
    masers become active, as a function of 2 model parameters.
    Contours are labelled with the maser frequency in GHz.  For each
    maser a single contour representing brightness temperature
    $10^{6}$~K is drawn, with no masers appearing above this threshold
    in the lower regions of the plots.  Top panel shows behaviour as
    \Tk\ and \Td\ are varied, when $\nH=10^{6.5}$~\cc\ and
    $\scdm=10^{12}$~\ccs.  Bottom panels show behaviour as \nH\ and
    \scdm\ are varied, when $\Td=150$~K for $\Tk=30$~K (left panel)
    and $\Tk=130$~K (right panel).  Points marked \lq A\rq\ to \lq
    G\rq\ illustrate different combinations of masers present
    simultaneously, as discussed in Section~\ref{sec:modelling} of the
    text.}
\label{fig:contours}
\end{figure*}

Figure~\ref{fig:contours} displays the results of model calculations
in which \Td, \Tk, \nH\ and \scdm\ are systematically varied.  Each
plot shows the result of varying two of these parameters, while the
other two are held constant, representing selected planes through the
parameter space.  Figure~\ref{fig:contours} displays a single contour
for each of the maser transitions listed in Table~\ref{tab:fluxes}
(6.6, 12.1, 85.5, 86.6, 107.0, 108.8, 156.6~GHz).  The contour value
is set at brightness temperature $10^6$~K.  Thus the area enclosed by
each contour represents combinations of model parameters which give
rise to masers at that frequency which are strong enough to be readily
observed.  Note that the exponential maser amplification leads to a
very rapid rise in maser brightness, with the 6.6-GHz maser attaining
brightness temperatures $>10^{11}$~K.  This style of plot is useful
for identifying regimes where different masers switch on
simultaneously.  Such plots can also be used to define limits on
parameter values, while recognising that they represent only a partial
exploration of the many dimensional phase space of the model.

The top panel of Fig.~\ref{fig:contours} shows the behaviour of the
model as dust and gas temperatures are varied, with the gas density
fixed at $\nH=10^{6.5}$~\cc\ and methanol specific column density
fixed at $\scdm=10^{12}$~\ccs.  Masers appear in the upper part of the
plot, but disappear if the dust temperature is too low (insufficient
pumping).  All the masers investigated here, with the exception of the
86.6-GHz line, are more readily excited at lower gas temperatures.
Under the conditions shown, the 6.6- and 12.1-GHz masers appear
together, requiring $\Td \ge 60$~K.  The 107.0-, 156.6- and 108.8-GHz
masers require $\Td \ge 125$~K, and progressively greater dust
temperatures in warmer gas.  The 85.5-GHz maser is confined to low gas
temperature, while the 86.6-GHz maser is favoured by higher gas
temperatures, with the two appearing together only for dust
temperatures $\Td > 200$~K and gas temperatures $\Tk \le 40$~K.  De
Buizer \etal\/ \shortcite{DPT00} measure dust temperatures in the
range of 100-200K for the mid-infrared sources associated with strong
6.6-GHz methanol masers, consistent with the general findings of the
modelling.

The lower panels of Fig.~\ref{fig:contours} show the behaviour of the
model as methanol column density and gas density are varied, with the
dust temperature fixed at $\Td=150$~K, for gas temperatures fixed at
$\Tk=30$ and 130~K in the left and right panels respectively.  Masers
appear above the threshold in the upper left parts of the plots, but
disappear if the methanol column density is too low (insufficient
amplification) or the gas density is too high (thermalisation of
populations).  The 6.6-GHz masers appear over the widest range of
conditions, and are accompanied by masers at 12.1~GHz except at the
lowest methanol column densities.  The 107.0-, 156.6- and 108.8-GHz
masers turn on in sequence as the methanol column density is
increased, but do not extend to such high gas density as the
centimetre wavelength masers. The 85.5-GHz maser is confined to low
gas temperature and intermediate gas densities, while the 86.6-GHz
maser has a somewhat complementary distribution, with the two
appearing together only for large methanol column density.

The majority of class~II methanol maser sources are
characterised by centimetre wavelength maser emission at 6.6 and/or
12.1~GHz, without maser counterparts at millimetre wavelengths.
Points marked \lq A\rq\ in Fig.~\ref{fig:contours} are examples of
model conditions producing masers above the threshold at 6.6~GHz only,
while points marked \lq B\rq\ produce 6.6- and 12.1-GHz masers
simultaneously.  The calculations suggest that these sources have dust
temperatures $60-150$~K, but do not place limits on their gas
temperatures.  The gas density is $\le 10^{8}$~\cc, with methanol
specific column density $\ge 10^{10.5}$~\ccs.  It is likely that the
$\sim50$ percent of these sources which are detectable at 6.6~GHz but
not at 12.1~GHz have lower values of methanol column density than the
sources exhibiting both masers.  The few sites known to exhibit masers
at 12.1~GHz only are probably at the high gas density limit of the
maser range.

In the following we assume that when masers at different frequencies
are detected at matching velocities, they are simultaneously excited.
Studies at high spatial resolution are required to confirm this by
establishing whether or not the different masers are positionally
coincident.  In comparison with the centimetre wavelength maser
sources, the 25 target sources which also have detectable masers at
107.0~GHz are likely to have slightly warmer dust temperatures, or to
have slightly greater methanol column density, but do not exceed gas
density $10^{7.5}$~\cc.  Points marked \lq C\rq\ in
Fig.~\ref{fig:contours} are examples of conditions producing masers
above the threshold at 107.0~GHz, but not at the other millimetre
wavelengths, as was observed for 18 of the 25 sources in
Table~\ref{tab:fluxes}.  The 3 sources with masers at 156.6~GHz (but
not at 85.5, 86.6 or 108.8~GHz) continue the trend to greater dust
temperature and/or methanol column density.  Points marked \lq D\rq\ 
in Fig.~\ref{fig:contours} are examples of this combination.  The
remaining 4 sources with 85.5-, 86.6- and/or 108.8-GHz masers
represent outliers in the population of maser sites.

We can estimate gas temperatures for these latter 4 sources: first
roughly from the regions in Fig.~\ref{fig:contours} with appropriate
combinations of masers, and subsequently with a more quantitative
approach to fitting model calculations to observations.  For example,
W3(OH) with masers at 6.6, 12.1, 107.0 and 86.6~GHz (but not 85.5,
108.8 or 156.6~GHz) is likely to be an example of a warm gas source,
with conditions resembling those of points \lq E\rq\ in
Fig.~\ref{fig:contours}.  This is confirmed by detailed modelling by
Sutton \etal\/ \shortcite{SSECMOG01} which obtained $\Tk=150$~K.  The
exceptional source 345.010+1.792 shows maser emission in all 7
transitions, and if we assume these to coincide then
Fig.~\ref{fig:contours} suggests a cool gas temperature, warm dust
temperature and large methanol column density, as exemplified by the
points marked \lq F\rq.  Detailed modelling by Cragg \etal\/
\shortcite{CSECGSD01} confirms this (Model B of that paper), but
predicts accompanying maser action at 94.5~GHz which was not detected.

The sources 328.808+0.633 and 9.621+0.196 exhibit masers at 6.6, 12.1,
107.0 and 85.5~GHz (but not at 86.6, 108.8 or 156.6~GHz), although in
both cases the 85.5-GHz maser is particularly weak.
Fig.~\ref{fig:contours} suggests a cool gas temperature, with
conditions resembling those of point \lq G\rq.  We have undertaken new
fitting of the model to observations, along the lines described in
Cragg \etal\/ \shortcite{CSECGSD01} and Sutton \etal\/
\shortcite{SSECMOG01}. Table~\ref{tab:models} displays a comparison
between the maser observations in 328.808+0.633 and 9.621+0.196 and
examples of cool gas models ($\Tk=20$~K) which partially account for
them.  The models illustrated in Table~\ref{tab:models} predict strong
masers at 6.6 and 12.1~GHz and weak masers at 85.5 and 107.0~GHz.  The
detection of very weak maser action at 85.5~GHz is consistent with
these models, as are the nondetections at 86.6, 108.8 and 156.6~GHz.
De Buizer \etal\/ \shortcite{DPT00} estimate the dust temperature in
the region of 328.808+0.633 to be approximately 125K, in good
agreement with the model in Table~\ref{tab:models}.  However, they
were unable to estimate a dust temperature for 9.621+0.196 due to the
non-detection of mid-infrared emission towards this source.  In both
cases the calculated 12.1-GHz maser intensity is considerably greater
than that observed.

In order to compare observed flux densities with calculated brightness
temperatures, we must make an assumption about the size of the maser
region.  For the examples shown in Table~\ref{tab:models}, the size
assumed is 60~milliarcseonds (mas) for 328.808+0.633 and 20~mas for
9.621+0.196, chosen to match the flux density observed at 6.6~GHz to
the brightness temperature calculated for these particular models.
The detailed study of milliarcseconds structures of methanol masers by
Minier \etal\/ \shortcite{MBC02} found that at both 6.6 and 12.2~GHz
the masers consist of a compact core (of the order of 0.5-3 mas in
size) surrounded by a more extended halo with dimensions approximately
an order of magnitude larger.  The sizes we have assumed are therefore
consistent with the dimensions of the halo component of the masers.
One of the sources examined in detail by Minier \etal\/
\shortcite{MBC02} was 9.621+0.196 at 12.2~GHz.  For the emission at
-0.5~\kms\/ they measured a halo size of $>50$~mas and a core of
2~mas.

The models presented above can account for the presence of different
maser line combinations in the different sources, but do not agree
quantitatively with all observations, as Table~\ref{tab:models}
illustrates.  This may be due to insufficient exploration of the
parameter space, or to deficiencies in the model.  While the maser
pumping is dominated by radiative processes, the 85.5- and 86.6-GHz
masers in particular are sensitive to the gas density and temperature,
and therefore to the collisional excitation rates, which are not
accurately known for methanol.  Some calculations with nonselective
collisional excitation rates by Sobolev \etal\/ \shortcite{SCG97b} and
Cragg \etal\/ \shortcite{CSECGSD01} illustrate how the maser
intensities at high gas densities can be affected by changes to the
collision model.  Until better collisional excitation rates become
available for methanol, the density estimates from maser modelling
will be subject to order-of-magnitude uncertainties.

The examples above illustrate how the combination of methanol maser
transitions observed in a particular source can be used to identify
plausible excitation conditions.  This approach is particularly
valuable for sources which exhibit masers in several transitions
simultaneously.  The majority of class~II methanol maser sources,
however, have been detected in only the 6.6- and/or 12.1-GHz
transitions.  While the model can identify minimum conditions of dust
temperature and methanol column density required to excite these
masers, together with an upper limit to the gas density, observations
of these two lines alone are not enough to fully characterise the
maser sites in these high-mass star forming regions.  Nevertheless,
the upper limits on nondetected lines provide a powerful constraint on
the conditions in such sources, particularly when data is available
for several lines sensitive to complementary factors.  As further
surveys are undertaken at a variety of methanol maser frequencies, it
should be possible to refine quantitative estimates of physical
conditions, both in the exceptional sources with masers at many
frequencies, and in the others which are more typical of high-mass
star formation regions.

\begin{table*}
\caption{Comparison of observations and modelling for class~II methanol 
  masers in 328.808+0.633 and 9.621+0.196.  The calculated flux density 
  $S_{\rm calc}$ depends on the size of the maser region, which is assumed
  to be 60~mas for 328.808+0.633 and 20~mas for 9.621+0.196, in order to 
  match the flux density observed $S_{\rm obs}$ to the calculated model 
  brightness temperature $T_{\rm calc}$ at 6.6~GHz.  The calculation 
  presented for 328.808+0.633 has model parameters $\Tk=20$~K, $\Td=100$~K, 
  $\nH=10^{6.3}$~\cc, $\scdm=10^{11.4}$~\ccs, while for 9.621+0.196 the 
  parameter values are $\Tk=20$~K, $\Td=125$~K, $\nH=10^{6.6}$~\cc, 
  $\scdm=10^{11.4}$~\ccs, and the remaining model parameters are as 
  described in Section~\ref{sec:modelling} of the text.}
\label{tab:models}
\centerline{
\begin{tabular}{lrrrrr} \hline
Source & Transition & Frequency & $S_{\rm obs}$ & $T_{\rm calc}$& $S_{\rm calc}$  \\ 
&& (GHz) & (Jy) & (K) & (Jy) \\ \hline
328.808+0.633 &&&&& \\
& $5_{1}-6_{0}\,A^{+}$  &   6.6 &    380  &  3.89E+09 & 380      \\
& $2_{0}-3_{-1}\,E$     &  12.1 &      7  &  4.48E+08 & 146      \\
& $6_{-2}-7_{-1}\,E$    &  85.5 &    1.6  &  8.98E+04 & 1.4      \\
& $7_{2}-6_{3}\,A^{-}$  &  86.6 & $<$2.4  & -1.83E+02 & $<0.1$   \\
& $3_{1}-4_{0}\,A^{+}$  & 107.0 &    5.5  &  1.20E+05 & 3.0      \\
& $0_{0}-1_{-1}\,E$     & 108.8 &   $<$6  & -4.39E+01 & $<0.1$   \\
& $2_{1}-3_{0}\,A^{+}$  & 156.6 &   $<$8  &  2.11E+02 & $<0.1$   \\
9.621+0.196 &&&&& \\
& $5_{1}-6_{0}\,A^{+}$  &   6.6 &    100 &  1.04E+10 & 100      \\
& $2_{0}-3_{-1}\,E$     &  12.1 &     10 &  1.85E+09 & 59.3     \\
& $6_{-2}-7_{-1}\,E$    &  85.5 &    1.2 &  6.53E+05 & 1.0      \\
& $7_{2}-6_{3}\,A^{-}$  &  86.6 & $<$2.3 & -1.81E+02 & $<0.1$   \\
& $3_{1}-4_{0}\,A^{+}$  & 107.0 &     22 &  5.44E+06 & 13.5     \\
& $0_{0}-1_{-1}\,E$     & 108.8 &   $<$3 &  3.82E+03 & $<0.1$   \\
& $2_{1}-3_{0}\,A^{+}$  & 156.6 &   $<$3 &  4.95E+05 & 2.6      \\
\hline 
\end{tabular}}

\end{table*}

\section{Conclusions}

We have undertaken a search for the 85.5- and 86.6-GHz transitions of
methanol towards 22 southern sources which exhibit 107-GHz methanol
masers.  These observations show that maser emission from both of
these transitions is rare.  We have examined the implications of these
findings in the context of the Sobolev \& Deguchi model for methanol
masers and find that our results are consistent with
\begin{enumerate}
\item The majority of class~II methanol maser sources, which show
  methanol maser emission at only 6.6 and 12.1~GHz, have dust
  temperature $\Td=60-150$~K, gas density $\nH \le 10^{8}$~\cc, and
  methanol specific column density $\scdm \ge 10^{10.5}$~\ccs.
\item The minority of sources which also show 107.0-GHz methanol maser
  emission have slightly higher \Td\ or slightly larger \scdm\ than
  those which don't.
\item The four known sources which also show 156.6-GHz methanol maser
  emission continue this trend to slightly larger \Td\ and \scdm.
\item The four known sources with 85.5- and 86.6-GHz methanol maser
  emission represent outliers in the population of maser sites, with
  85.5-GHz maser emission characteristic of low gas temperature
  ($\Tk=20$~K), 86.6-GHz maser emission characteristic of high gas
  temperature ($\Tk=150$~K), and both present simultaneously only if
  $\Td > 200$~K or $\scdm > 10^{12.5}$~\ccs.
\end{enumerate}

A comparison of the 85.5- and 86.6-GHz methanol maser spectra observed
towards 345.010+1.792 with published and unpublished spectra from
earlier epochs shows clear variability, particularly for the 86.6-GHz
transitions.  In combination with the narrow line width of the
emission and the offset of the 85.5- and 86.6-GHz emission from the
velocity of the thermal emission for other molecular species in
345.010+1.792 this presents conclusive evidence that these transitions
are masing in this exceptional source.

\section*{Acknowledgements}

We would like to thank Paolo Calisse and John Shobbrook for their
assistance with the observations and the University of NSW and the
ATNF for their assistance in arranging observing support, in
particular Lucyna Kedziora-Chudczer, Michael Burton and Bob Sault.
The Australia Telescope is funded by the Commonwealth of Australia for
operation of a National Facility managed by CSIRO.  This research has
made use of NASA's Astrophysics Data System Abstract Service.  PDG and
DMC thank the Victorian Partnership for Advanced Computing for
financial support.

\end{document}